\documentclass[
superscriptaddress,
showkeys,
showpacs,
amsmath,amssymb,
pre,
twocolumn,
]{revtex4-2}

\usepackage{lineno}
\usepackage[colorlinks,linkcolor=red,citecolor=blue,urlcolor=blue]{hyperref}
\usepackage{amsfonts}
\usepackage{amssymb}
\usepackage{graphicx}
\usepackage{epstopdf}
\usepackage{pgfplots}
\pgfplotsset{compat=1.14}
\usepackage{bm}
\usepackage{soul}
\usepackage{comment}

\usepackage[draft]{todonotes}

\begin{document}
	\title{Probing double distribution function models in the lattice Boltzmann method for highly compressible flows}
	\author{S. A. Hosseini}
	\affiliation{Department of Mechanical and Process Engineering, ETH Zurich, 8092 Zurich, Switzerland}%
 	\author{A. Bhadauria}
	\affiliation{Department of Mechanical and Process Engineering, ETH Zurich, 8092 Zurich, Switzerland}%
	\author{I. V. Karlin}\thanks{Corresponding author}
 \email{ikarlin@ethz.ch}
	\affiliation{Department of Mechanical and Process Engineering, ETH Zurich, 8092 Zurich, Switzerland}%
	\date{\today}
	\begin{abstract}
		{The double distribution function approach is an efficient route towards extension of kinetic solvers to compressible flows. With a number of realizations available, an overview and comparative study in the context of high speed compressible flows is presented. We discuss the different variants of the energy partition, analyses of hydrodynamic limits and a numerical study of accuracy and performance with the particles on demand realization. Out of three considered energy partition strategies, it is shown that the non-translational energy split requires a higher-order quadrature for proper recovery of the Navier--Stokes--Fourier equations. The internal energy split on the other hand, while recovering the correct hydrodynamic limit with fourth-order quadrature, comes with a non-local --both in space and time-- source term which contributes to higher computational cost and memory overhead. Based on our analysis, the total energy split demonstrates the optimal overall performance. 
  }
	\end{abstract}
	\keywords{}
	\maketitle
\section{Introduction}
Extension of the lattice Boltzmann method (LBM), and other discrete velocity methods for the Boltzmann equation, to compressible flow simulations has been a topic of interest over the past decade~\cite{hosseini2024lattice}. A number of different strategies have been devised to extend the original isothermal weakly compressible LBM to compressible flows with energy balance. One straightforward approach is the use of larger lattices allowing the model to properly recover more moments of the equilibrium distribution function -necessary for the energy balance up to Navier--Stokes--Fourier level \cite{philippi2006continuous,shan2006kinetic,siebert2008lattice,frapolli2014multispeed}. This strategy, as it has been observed over the past few years, has a number of limitations such as an increase in non-locality of the streaming operator, larger computational load and memory footprint. It also has a limited range of operation in terms of the temperature. 

An alternative, allowing the use of smaller lattices, is the so-called double distribution function approach \cite{he1998novel,guo2007thermal,karlin2013consistent,asinari2010quasiequilibrium,hosseini2023lattice} inspired by Rykov's original work \cite{rykov1975model}. In this approach, the distribution function is split into two reduced distribution functions where the second one carries a form of energy. This idea has become a popular approach in discrete kinetic theory of gases, in combination with the Bhatnagar--Gross--Krook (BGK) and similar collision operators in order to extend isothermal solvers to compressible flows. While in Rykov's original model, the second distribution function was intended to solely transport internal energy due to non-translational degrees of freedom for polyatomic molecules and did not make any assumptions as to equilibrium between internal and translational degrees of freedom, in the context of discrete solvers the choice of the variable transported by the second distribution function is not unique and usually assumes equipartition, though recent proposals with non-equilibrium temperatures can be found \cite{kolluru2020extended}. Different choices of variables have been used and presented over the years. The aim of the present contribution is to discuss the non-uniqueness in the choice of the variable carried by the second population by looking at and analyzing the effects of each one of the popular choices in the literature, with a focus on applications involving larger Mach numbers. The focus of the discussion being on the double distribution function approach and effect of the choice of variable carried by the second distribution we will only consider BGK collision operators. Nonetheless, the results can readily be extended to variable Prandtl numbers and/or independent bulk viscosity via introduction of quasi-equilibrium states~\cite{gorban1994general,ansumali2007quasi}, for instance. 
The article will begin with a brief overview of the double distribution function formalism and then discuss the consequences of three major choices, namely non-translational internal energy, internal energy and total energy, all in the context of a single relaxation time BGK collision operator. Finally numerical simulation results, conducted using a particles-on-demand realization~\cite{dorschner2018particles}, probing all hydrodynamic level properties will be presented and discussed.
\section{Background and theory}
In the context of the present contribution we are interested in the Boltzmann equation with a BGK approximation for the collision operator~\cite{bhatnagar1954model}, i.e.
\begin{equation}\label{eq:boltz_eq}
    \partial_t f + \bm{v}\cdot\bm{\nabla} f = \frac{1}{\tau}\left(f - f^{\rm eq}\right).
\end{equation}
Here, $f$ represents the first reduced probability distribution function with $\bm{v}$ the particle velocity, $\bm{x}$ the position in physical space and $t$ time. The definition of the first reduced distribution function, for molecules endowed with internal degrees of freedom follows that of Rykov in~\cite{rykov1975model}.
The reduced distribution function $f$ is subject to the balance equation \eqref{eq:boltz_eq}, with the equilibrium defined as,
\begin{equation}\label{eq:f_eq}
	f^{\rm eq} = \frac{\rho }{{(2\pi r T )}^{D/2}} \exp{\left(-\frac{{(\bm{v}-\bm{u})}^2}{2rT}\right)},
\end{equation}
where $T$ is the temperature, $\rho$ density, $\bm{u}$ velocity and $r$ the gas constant. 
Moreover, $\tau$ is the relaxation time, tied to the fluid dynamic viscosity, $\mu$, as,
\begin{equation}\label{eq:kin_visc}
    \tau = \frac{\mu}{p},
\end{equation}
where $p=\rho r T$ is the thermodynamic pressure. While the zeroth- and first-order conserved moments are computed with the first reduced distribution function,
\begin{subequations}
\begin{align}
	   \int_{\mathbb{R}^D} f d\bm{v} &= \rho,\\
	   \int_{\mathbb{R}^D} \bm{v} f d\bm{v} &= \rho\bm{u},
\end{align}
\end{subequations}
the second reduced distribution function carries a form of energy. 
Let us introduce the kinetic energy,
\begin{equation}
	\label{eq:kin_energy}
	K=\frac{1}{2}\rho\bm{u}^2,
\end{equation} and the internal energy,
\begin{equation}\label{eq:int_energy}
    U = \rho\int_0^T c_v dT,
\end{equation}
where $c_v$ is specific heat at constant volume. The total energy $E$ is
\begin{equation}
	E = U + K.
\end{equation}
Furthermore, let $U'$ be the part of the internal energy associated with the non-translational degrees of freedom,
\begin{equation}
  U' = U-\rho\frac{DrT}{2}.
\end{equation}
The second distribution function can now be defined with respect to the total energy in several ways,
\begin{subequations}\label{eq:Es}
\begin{align}
	   E &= \int_{\mathbb{R}^D} g d\bm{v}, \label{eq:total_energy_g}\\
	   E &= \int_{\mathbb{R}^D} g  d\bm{v}+ K, \label{eq:int_energy_g}\\
      E &= \int_{\mathbb{R}^D} \left(g + \frac{\bm{v}^2}{2}f\right) d\bm{v}. \label{eq:nt_energy_g}
\end{align}
\end{subequations}
Note that the approach of Eq.~\eqref{eq:total_energy_g} was first proposed in \cite{guo2007thermal,karlin2013consistent} in the context of the LBM while Eqs.~\eqref{eq:int_energy_g} and \eqref{eq:nt_energy_g} were discussed in \cite{he1998novel} and \cite{frapolli2014multispeed}, respectively. The model discussed here is slightly different from \cite{he1998novel} in that it also accounts for internal degrees of freedom. For the sake of readability, in the remainder of the paper we will refer to Eqs.\eqref{eq:total_energy_g}, \eqref{eq:int_energy_g} and \eqref{eq:nt_energy_g}, respectively as the total, internal and non-translational energy splits. It is interesting to note that, while replacing the temperature $T$ in the Maxwell--Boltzmann equilibrium \eqref{eq:f_eq} with internal energy through Eq.~\eqref{eq:int_energy}, the reduced $f$-equilibrium becomes dependent on the second $g$-distribution function which carries a part or all of the internal energy. 
The following kinetic equations for the different realizations \eqref{eq:Es} of the second reduced distribution functions can be derived,
\begin{subequations}
\begin{align}
	   \partial_t g + \bm{v}\cdot\bm{\nabla} g &= -\frac{1}{\tau}\left(g - g^{\rm eq}\right),\\
	   \partial_t g + \bm{v}\cdot\bm{\nabla} g &= -\frac{1}{\tau}\left(g - g^{\rm eq}\right)  - f(\bm{v}-\bm{u})\cdot\left(\partial_t \bm{u} + \bm{v}\cdot\bm{\nabla}\bm{u}\right),\label{eq:int_g_evolution}\\
      \partial_t g + \bm{v}\cdot\bm{\nabla} g &= -\frac{1}{\tau}\left(g - g^{\rm eq}\right),
\end{align}
\end{subequations}
where the reduced equilibrium distributions $g^{\rm eq}$ are respectively defined as,
\begin{subequations}
\begin{align}
g^{\rm eq} &= \left(\frac{1}{\rho}U' + \frac{{\bm{v}}^2}{2}\right) f^{\rm eq},\\
g^{\rm eq} &= \left(\frac{1}{\rho}U' + \frac{{(\bm{v}-\bm{u})}^2}{2}\right) f^{\rm eq},\\
g^{\rm eq} &= \frac{1}{\rho}U' f^{\rm eq}.
\end{align}
\end{subequations}
Details of the multi-scale Chapman-Enskog analysis are presented in the Appendix \ref{app:CE}.
All realizations \eqref{eq:Es} lead to the Navier--Stokes equations for the momentum balance,
\begin{equation}
    \partial_t\rho \bm{u} + \bm{\nabla}\cdot\rho\bm{u}\otimes\bm{u} + \bm{\nabla}p - \bm{\nabla}\cdot\bm{T}_{\rm NS}  = 0,
\end{equation}
where the viscous stress tensor $\bm{T}_{\rm NS}$ is defined as,
\begin{equation}
    \bm{T}_{\rm NS} = \mu\left[\bm{\nabla}\bm{u} + \bm{\nabla}\bm{u}^\dagger - \frac{2}{D}\bm{\nabla}\cdot\bm{u}\bm{I}\right] + \eta \bm{\nabla}\cdot\bm{u}\bm{I}.
\end{equation}
Here we have introduced the bulk viscosity, $\eta$.
A closer look at the results show that all double distribution realizations maintain consistent bulk viscosity, i.e.
\begin{equation}\label{eq:bulk_visc}
    \eta = \left(\frac{2}{D}-\frac{r}{c_v}\right)\mu.
\end{equation}
Finally, at the energy level all realizations recover the following,
\begin{equation}
    \partial_t E + \bm{\nabla}\cdot\left(E + p\right)\bm{u} + \bm{\nabla}\cdot\bm{q}_{\rm NS} = 0,
\end{equation}
with,
\begin{equation}
    \bm{q}_{\rm NS} = -\lambda\bm{\nabla}T - \bm{u}\cdot\bm{T}_{\rm NS},
\end{equation}
where the first term is the Fourier heat flux and the second viscous heating. Note that the thermal conductivity $\lambda$ is tied to the relaxation time as,
\begin{equation}\label{eq:therm_cond}
    \lambda = \tau p (c_v + r),
\end{equation}
leading to fixed Prandtl number
\begin{equation}
    {\rm Pr} = \frac{(c_v + r) p\tau}{(c_v + r) p\tau} = 1,
\end{equation}
and adiabatic exponent
\begin{equation}\label{eq:spec_heat}
    \gamma = \frac{c_v + r}{c_v}.
\end{equation}
In this study, we focus on the impact of various partitions of the energy \eqref{eq:Es} and thus use the simplest single relaxation time kinetic model for the sake of presentation. This certainly restricts the Prandtl number.
Nevertheless, all energy partitions \eqref{eq:Es} discussed in the present study can be readily extended to a tunable Prandtl number, by introducing an intermediary state of relaxation via the quasi-equilibrium approach as discussed in \cite{gorban1994general}.
These intermediary states are constructed via minimization of entropy under constraints defined by a set of select higher-order moments. The choice of additional constraints in the set as compared to the conserved moments, determines the fluxes with independent relaxation rates.
Specific discussions on the quasi-equilibrium approach for independent control over bulk viscosity and Prandtl number can be found in  \cite{gorban1994general,ansumali2007quasi,asinari2010quasiequilibrium,hosseini2023lattice}.

Now that the hydrodynamic limits of each realization has been clarified, let us go over minimum requirements for the discrete solver in each case; In discrete velocity methods such as LBM, the minimum number of degrees of freedom --i.e. discrete velocities-- is determined by the number of moments of the equilibrium distribution function that need to be satisfied to recover the target hydrodynamic limit.

For instance, in the full energy model, the first reduced distribution function $f$ only needs to properly recover continuity and momentum balance equations to the Navier--Stokes order in the Chapman--Enskog expansion, which involves equilibrium moments up to order three, while the second distribution function $g$ requires accuracy of the equilibrium moments up to second order. 
This can be achieved either through a fourth-order quadrature, i.e. D$D$Q$4^D$ lattice, or a third order quadrature, i.e. D$D$Q$3^D$ lattice, with a correction term  for the third-order moments~\cite{li2007coupled,prasianakis2007lattice,hosseini2020compressibility}.
For the internal energy model, the constraints on $f$ and $g$ remain the same, as the energy balance equation relies on a combination of the zeroth-order moments of both.
The last realization, i.e. the non-translational splitting, defines energy as the sum of the second-order moment of $f$ and zeroth-order moment of $g$, which brings in an additional constraint on the fourth-order moment of the $f$ distribution function, 
\begin{equation*}
    \int_{\mathbb{R}^D} \bm{v}\otimes\bm{v}\otimes\bm{v}\otimes\bm{v} f d\bm{v} = \sum_{i=1}^{Q} \bm{c}_i\otimes\bm{c}_i\otimes\bm{c}_i\otimes\bm{c}_i f_i,
\end{equation*}
and which increases the minimum lattice size to D$D$Q$5^D$.
The present study will use fourth- and fifth-order quadrature-based lattices. Details of these lattices are given in Table~\ref{tab:lattices}.
\begin{widetext}
	\begin{center}
		\begin{table}[h!]
			\begin{tabular}{ |p{2cm}||p{1cm}|p{5cm}|p{6cm}|p{2cm}|  }
				\hline
				Quadrature & $r T_L$ & $c_{i\alpha}$ & $w_{i\alpha}$ & 2-D lattice\\
				\hline
				D1Q3  & 1 & $\{0, \pm\sqrt{3}\}$ & $\{2/3, 1/3\}$ & D2Q9\\
				\hline
				D1Q4 & 1 & $\{\pm\sqrt{3-\sqrt{6}}, \pm\sqrt{3+\sqrt{6}}\}$ & $\{(3+\sqrt{6})/12, (3-\sqrt{6})/12\}$ & D2Q16\\
				\hline
				D1Q5 & 1 & $\{0, \pm\sqrt{5-\sqrt{10}}, \pm\sqrt{5+\sqrt{10}}\}$ & $\{8/15, (7+2\sqrt{10})/60, (7-2\sqrt{10})/60\}$ & D2Q25\\
				\hline
			\end{tabular}
			\caption{Details of quadratures used for simulations.}
			\label{tab:lattices}
		\end{table}
	\end{center}
\end{widetext}
The above discussion establishes directly computational cost and memory footprint of each realization; This analysis has to be complemented with one further consideration which is the presence of source terms in Eq.~\eqref{eq:int_g_evolution} involving space- and time-derivatives of $\bm{u}$ which brings in non-negligible computational cost, along with additional memory footprint due to the time-derivative. Finally, the presence of such non-local contributions, in the case of finite-differences approximations, as used for instance in \cite{he1998novel}, contributes to non-conservation issue which can become detrimental in high Mach number flows simulations. An overview of the properties of each splitting approaches is presented in Table \ref{tab:properties}.
\begin{widetext}
\begin{center}
\begin{table}[h!]
\begin{tabular}{ |p{6cm}||p{2cm}|p{2cm}|p{2cm}|  }
\hline
 Energy split & Eq.~\eqref{eq:total_energy_g} &  Eq.~\eqref{eq:int_energy_g} & Eq.~\eqref{eq:nt_energy_g}\\
 \hline
 Kinematic viscosity  & Eq.~\eqref{eq:kin_visc} & Eq.~\eqref{eq:kin_visc} & Eq.~\eqref{eq:kin_visc}\\
 \hline
 Bulk viscosity & Eq.~\eqref{eq:bulk_visc} & Eq.~\eqref{eq:bulk_visc} & Eq.~\eqref{eq:bulk_visc}\\
 Thermal conductivity & Eq.~\eqref{eq:therm_cond} & Eq.~\eqref{eq:therm_cond} & Eq.~\eqref{eq:therm_cond}\\
 Specific heat ratio & Eq.~\eqref{eq:spec_heat} & Eq.~\eqref{eq:spec_heat} & Eq.~\eqref{eq:spec_heat}\\
 Minimum order of quadrature for NSF &  four  &  four & five\\
 Non-local source terms & No  & Yes   & No\\
 \hline
\end{tabular}
\caption{Overview of properties of different splitting strategies.}
\label{tab:properties}
\end{table}
\end{center}
\end{widetext}
In the next section, we will probe points discussed here through numerical simulations.
\section{Numerical applications}
In this section, simulation are conducted to verify theoretical results derived in the previous sections. First the numerical scheme is introduced and then simulation results are presented and discussed. For the sake of readability all variables will be presented in lattice units, i.e. normalized by grid and time-steps sizes, $\delta x$ and $\delta t$; Velocities are normalized by $\delta x/\delta t$ and temperatures $rT$ by $\delta x^2/\delta t^2$.
\subsection{Discrete velocity solver for highly compressible flows: PonD}
As discussed in the introduction, given that final target of the present contribution is application to highly compressible flows, we use a numerical model developed in our group, \cite{dorschner2018particles}, shown to be particularly well-adapted to such flows, i.e. the particles on Demand (PonD) method. To minimize errors in higher-order moments of the distribution function not supported by the lattice quadrature, in the PonD method \cite{dorschner2018particles}, the reference frame is chosen via the local macroscopic quantities, i.e. fluid speed and temperature
\begin{equation}
    rT^{\lambda} = rT,\ \bm{u}^\lambda = \bm{u}, 
\end{equation}
and the particle velocity is locally defined as, 
\begin{equation}\bm{C}_i(rT^\lambda, \bm{u}^\lambda) = \sqrt{rT^\lambda} \bm{c}_i + \bm{u}^\lambda.\end{equation}
In so doing, equilibrium distribution functions are accurately evaluated and only depend on the local density
\begin{equation}
    f_i^{eq} = \rho w_i
\end{equation}
where $w_i$ are weights of the Gauss-Hermite quadrature \cite{frapolli2015}. Different from approaches such as \citep{kauf2011multi,kollermeier2013hyperbolic} where an adaptive reference-based Boltzmann equation is solved \emph{globally} in the domain bringing in non-commutativity of the discrete velocities with the space and time-derivatives and resulting in additional terms in the Boltzmann equation involving derivatives of the reference frame velocity and temperature, in PonD at every point in space and time $\bm{x}$ and $t$ the Boltzmann equation is solved in the fixed reference frame of that point, here denoted as $\bar{\lambda}$ for the sake of readability, leading to the following local evolution equation:
\begin{equation}\label{eq:Boltzmann_loc_eq}
    D_t^{\bar{\lambda}} \mathcal{M}_{\lambda(\bm{x},t)}^{\bar{\lambda}}\{f_i^{\lambda(\bm{x},t)}, g_i^{\lambda(\bm{x},t)}\}  = \mathcal{M}_{\lambda(\bm{x},t)}^{\bar{\lambda}} \Omega_{\{f_i^{\lambda(\bm{x},t)}, g_i^{\lambda(\bm{x},t)}\}},
\end{equation}
where $\mathcal{M}_{\lambda(\bm{x},t)}^{\bar{\lambda}} $ is the reference frame transformation operator.
\subsubsection{Frame transformation}
In the previous section we have introduced the reference frame transformation, $\mathcal{M}_{\lambda(\bm{x},t)}^{\bar{\lambda}}$, which allowing information to be transformed from one reference frame to other is an essential component of PonD. The main statement, allowing for such an operation is that discrete distribution functions can be equivalently written as the set of moment correctly matched by the quadrature and that these moment are frame-invariant, i.e.
\begin{align}
    \sum_{i=0}^{Q-1} f_i C_{ix}^p(rT^\lambda,u_x^\lambda) C_{iy}^q(rT^\lambda,u_y^\lambda) C_{iz}^r(rT^\lambda,u_z^\lambda) &= M^{\lambda}_{x^p y^q z^r},
    \label{eq:genericMomentUptoO3}
\end{align}
This approach, also referred to as the moment matching method was developed and used in \cite{dorschner2018particles}. Here, in an effort to further control errors in higher-order moments not supported by the lattice during transformation, we make use of a reduced order Grad expansion~\cite{grad_kinetic_1949}, i.e. a regularized reconstruction of discrete populations~\cite{Zipunova_Reg2021,kallikounis2022particles,sawantPonD_2022}. The distribution functions in a reference frame $\bar{\lambda}$ can be computed from moment in a reference frame $\lambda$ as,
\begin{align}
    &f_i^{\bar{\lambda}} = w_i \sum_{n=0}^{N} \frac{1}{n!rT_L} a^{\bar{\lambda}}_{(n)}:H_i^{(n)},
    \label{eq:gradGeneralTransform}
\end{align}
where $a^\lambda_{(n)}$ is the rank $n$ tensor of Hermite polynomials of the corresponding order in the reference frame $\lambda$ and $H_i^{(n)}$ the tensor of Hermite coefficients of the same order, $:$ the Frobenius inner product and $N$ the order of the Grad expansion. The Grad coefficients in reference frame $\bar{\lambda}$ can be computed from moments in the reference frame $\lambda$ as,
\begin{equation}\label{eq_a0lp}
    a_0^{\bar{\lambda}} = M^\lambda_0,
\end{equation}
\begin{equation}
    a_1^{\bar{\lambda}} = \frac{1}{\sqrt{T^{\bar{\lambda}}/T_L}}\left(M^\lambda_1 - M_0^\lambda u^{\bar{\lambda}}\right),
\end{equation}\label{eq_a1lp}
\begin{equation}
    a_{2}^{\bar{\lambda}} = \frac{T_L}{T^{\bar{\lambda}}}\left(M^\lambda_2 - M_0^\lambda rT^{\bar{\lambda}} \bm{I} - \sqrt{\frac{T^{\bar{\lambda}}}{T_L}}\overline{u^{\bar{\lambda}}\otimes a_1^{\bar{\lambda}}}  - M_0^\lambda u^{\bar{\lambda}}\otimes u^{\bar{\lambda}}\right),
\end{equation}
and
\begin{multline}\label{eq_a2lp}
    a_{3}^{\bar{\lambda}} = \frac{1}{{\left(T^{\bar{\lambda}}/T_L\right)}^{3/2}}\left(M^\lambda_3 \right. \\ \left. - \frac{T^{\bar{\lambda}}}{T_L} \overline{u^{\bar{\lambda}}\otimes\left(M_0^\lambda T_L\bm{I} + a_2^{\bar{\lambda}}\right)} -  T^{\bar{\lambda}}\sqrt{\frac{T^{\bar{\lambda}}}{T_L}}  \overline{a_1^{\bar{\lambda}}\otimes \bm{I}} \right. \\ \left. - \sqrt{\frac{T^{\bar{\lambda}}}{T_L}} \overline{a_1^{\bar{\lambda}}\otimes u^{\bar{\lambda}} \otimes u^{\bar{\lambda}}}  - M_0^{\lambda} u^{\bar{\lambda}} \otimes u^{\bar{\lambda}} \otimes u^{\bar{\lambda}}\right).
\end{multline}
Here Eqs.~\eqref{eq:gradGeneralTransform} through \eqref{eq_a2lp} define the reference frame transformation operaor.
\subsubsection{Streaming}
While different space/time discretization approaches have been developed by our group in recent years, see for instance~\cite{ji2024eulerian,kallikounis2022particles}, we will use the semi-Lagrangian discretization here. Propagation occurs along the co-moving particle velocities which are fully adaptive, and therefore no longer space filling,
\begin{equation}
    \{f_i, g_i\}^{\lambda}(\bm{x}+ \bm{C}_i\delta t, t+\delta t) = \{f_i^*, g_i^*\}^{\lambda}(\bm{x}, t),
\end{equation}
where $\{f_i^*, g_i^*\}$ are post-collision distribution functions. As such, the streaming step is supplemented with a reconstruction step to get the discrete distribution functions on the grid points. Each of these populations exist in their local reference frames, which are distinct from one another and also distinct from the destination frame. We can write for a generalized $p$-sized interpolation kernel $W(\bm{x})$ :
    \begin{align}
         \{f_i, g_i\}^{\bar{\lambda}}(\bm{x},t) = \sum_{s=0}^{p-1} W(\bm{x}-\bm{x}_s)  \mathcal{M}_{\lambda_s}^{\bar{\lambda}}  \{f_i, g_i\}^{\lambda_s}(\bm{x}_s,t)
        \label{eq:propagationInterpolated}
    \end{align} 
where $\bm{x}_s$ is the position of a node inside the interpolation kernel and $\lambda_s$ is the local reference frame at this node. In the context of the present work the reconstruction step is conducted through a 2nd-order Lagrangian interpolation stencil.
\subsection{Probing hydrodynamic limits}
\subsubsection{Speed of sound: Effect of specific heats ratio}
\begin{figure}
    \centering
    \includegraphics[width=0.9\linewidth]{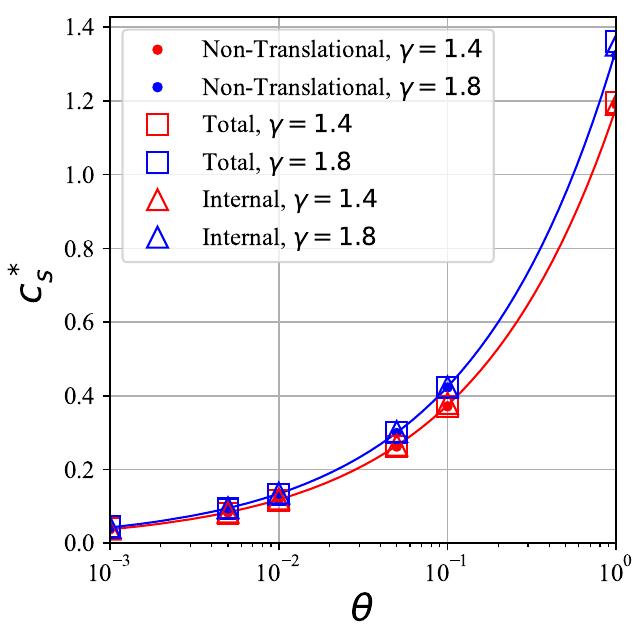}
    \caption{ Speed of sound on the lattice plotted against different values of non-dimensional temperature. Red line indicates results for $\gamma=1.4$ and blue line $\gamma=1.8$. }
    \label{fig:sound_speed}
\end{figure}

As a first step and to validate the dispersion properties of normal modes, we look at the temperature-dependence and effect of $c_v$ on the speed of sound. A freely travelling pressure front is simulated to that effect. A quasi-one-dimensional domain $ L_x \times 1$ (with $L_x = 800$) is separated into two parts with a pressure difference of $\Delta p=10^{-4}$ and a uniform initial temperature $T_0$ and velocity $\bm{u}_0=0$. The speed of sound is computed by tracking the position of the shock front over time and compared with the theoretical value $c_s(T) = \sqrt{\gamma r T}$. The simulations are performed with two different specific heat ratios, i.e. $\gamma = 1.4$ and $1.8$, in a wide range of temperature. From figure~\ref{fig:sound_speed} we observe that all energy splitting strategies, as expected, can correctly capture the speed of sound at the considered range of temperatures. Note that temperatures are reported in non-dimensional form, i.e. {$\theta = T/T_L$} and speed of sound is in lattice units.
\subsubsection{Measuring dissipation rates}
Next we look into the dissipation rates of the three hydrodynamic modes, i.e. shear, normal and entropic. From the Chapman-Enskog analysis, the kinematic shear viscosity $\nu=\mu/\rho$ and bulk viscosity $\eta$ in all splittings are related to the relaxation coefficient $\tau$ as,
\begin{subequations}
\begin{align}
	   \nu &= \frac{\mu}{\rho} = \tau r T,\\
	   \frac{\eta}{\rho} &= \left(\frac{2}{D}-\frac{1}{c_v}\right)\tau r T,\\
      \alpha &= \frac{\lambda}{(c_v+r) \rho} = \tau r T.
\end{align}
\end{subequations}
To measure effective dissipation rates we conducted three set of simulations.\\
\par
\begin{figure}
    \centering
    \includegraphics[width=0.9\linewidth]{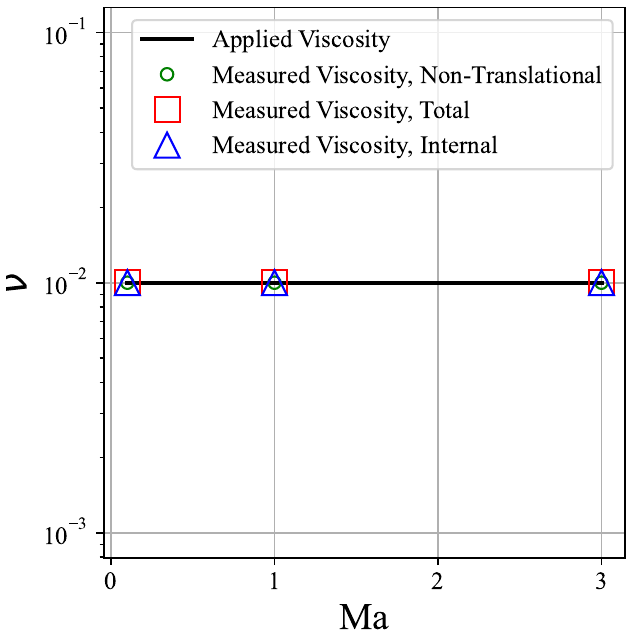}
    \caption{Measured effective kinematic viscosity as a function of Mach number for different energy splits.}
    \label{ShearVisc}
\end{figure}

 First, to measure the kinematic shear viscosity a plane shear wave is simulated. The wave is introduced via a small sinusoidal perturbation added to the initial velocity field. The initial conditions of the flows are
\begin{equation}
    \rho = \rho_0, T = T_0, u_x = u_0, u_y = A \sin\left(2\pi x/L_x\right).
\end{equation}
where the initial density and temperature is $ \left( \rho _0,T_0 \right) = \left( 1.0,1.0 \right) $, the perturbation amplitude {$A=1e-6$} and $u_0=0$. The simulation domain is $ L_x \times 1$ (with $L_x = 1600$). The shear viscosity is measured by monitoring the time evolution of the maximum velocity and fitting an exponential function to it, i.e.
\begin{equation}
    u_x^{\rm max}(t) \propto \exp{\left(-\frac{4\pi^2\nu}{L_x^2}t\right)}.
\end{equation}
Note that convergence studies were conducted as preliminary runs and all results reported here correspond to converged simulations in time and space. The results from simulations for different Mach numbers as obtained using the different splitting strategies are displayed in Fig.~\eqref{ShearVisc}.
The results clearly show the invariance of the effective kinematic viscosity with respect to the Mach number.\\
\par
\begin{figure}
    \centering
    \includegraphics[width=0.9\linewidth]{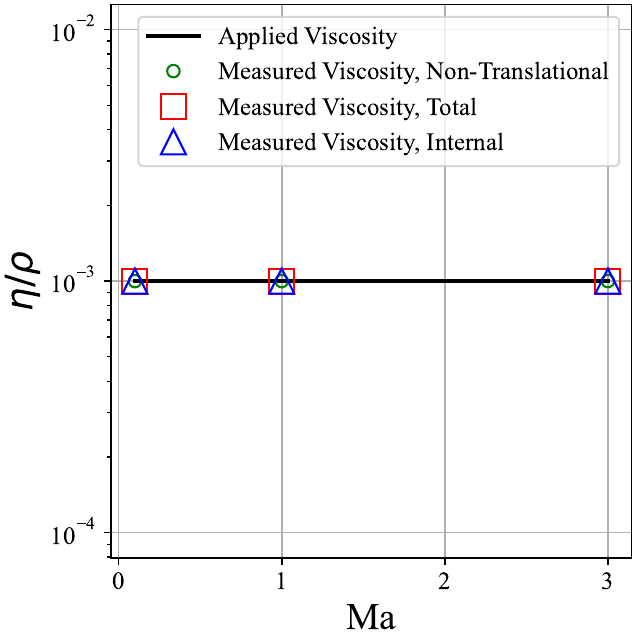}
    \caption{Measured effective bulk viscosity as a function of Mach number for different energy splits.}
    \label{BulkVisc}
\end{figure}

 The bulk viscosity is measured through the decay rate of sound waves. To that end a small perturbation is added to the uniform initial density field \citep{dellar2001bulk,hosseini2020compressibility}. Note that the smallness of the perturbation ensures that the study is in the linear regime, excluding effects such as non-linear steepening. For a discussion on non-linear acoustics readers are referred to studies such as \citep{buick2000lattice}. In the linear regime it is readily shown that wave dynamics are governed by the linear lossy wave equation, see \citep{lamb1924hydrodynamics}. The flow is initialized as
\begin{equation}
    \rho = \rho_0 + A \sin\left(2\pi x/L_x\right), T = T_0, u_x = 0, u_y = 0,
\end{equation}
where $\rho_0 = 1.0$ with initial amplitude $A=1e-6$ and the density perturbation is $\rho' = \rho - \rho_0 $. The initial temperature is $T_0 = 1.0$. The decay of energy $E(t) = u_x^2+u_y^2-u_0^2+c_s^2\rho '^2$ over time is supposed to fit an exponential function depending upon an effective viscosity $\nu_e$ which is a combination of kinematic shear and bulk viscosity
\begin{equation}
    \nu_e = \frac{4}{3} \nu + \frac{\eta}{\rho},
\end{equation}
defined by~\cite{dellar2001bulk} as
\begin{equation}
    E(t) \propto \exp{\left(-\frac{4\pi^2\nu_e}{L_x^2}t\right)}.
\end{equation}
The resolution is identical to the previous test case. Measured bulk viscosities for the different energy splits are shown in Fig.~\eqref{BulkVisc}.
In agreement with the multi-scale analysis, all three schemes recover the predicted bulk viscosity and are invariant with respect to the Mach number.\\
\par
\begin{figure}
    \centering
    \includegraphics[width=0.9\linewidth]{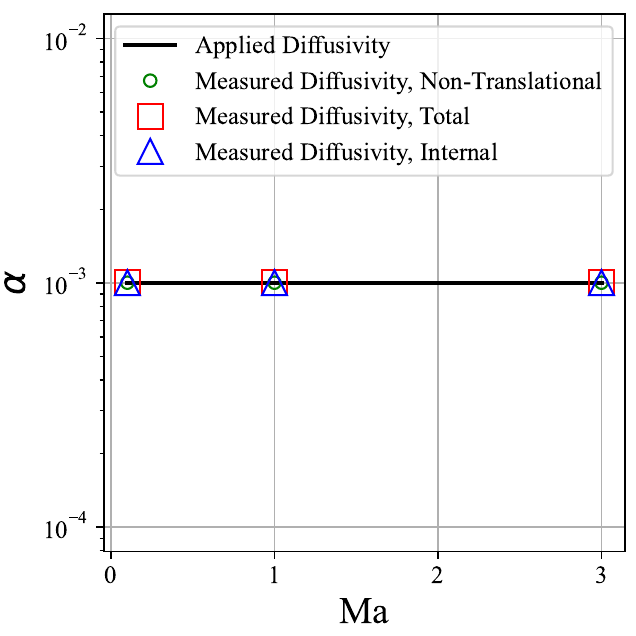}
    \caption{Measured effective thermal diffusivity as a function of Mach number for different energy splits.}
    \label{thermal_diffusivity}
\end{figure}

 To assess the thermal diffusivity $\alpha$, a different type of perturbation is introduced into the initial state of the system,
\begin{equation}
    \rho = \rho_0 + A sin\left(2\pi x/L_x\right), T = \rho_0 T_0/\rho, u_x = 0, u_y = u_0,
\end{equation}
where the initial density and temperature are $ \left( \rho _0,T_0 \right) = \left( 1.0,1.0 \right) $, the perturbation amplitude $A=1e-6$. The thermal diffusivity is measured through the time evolution of maximum temperature difference $T' = T_0 - T$,
\begin{equation}
    T'(t) \propto \exp{\left(-\frac{4\pi^2\alpha}{L_x^2}t\right)}.
\end{equation}
Figure ~\ref{thermal_diffusivity} plots the measured thermal diffusivity at different Mach numbers compared with the intended values $\alpha=10^{-3}$.
\subsubsection{Measuring viscous heating}
\begin{figure}
    \centering
    \includegraphics[width=0.9\linewidth]{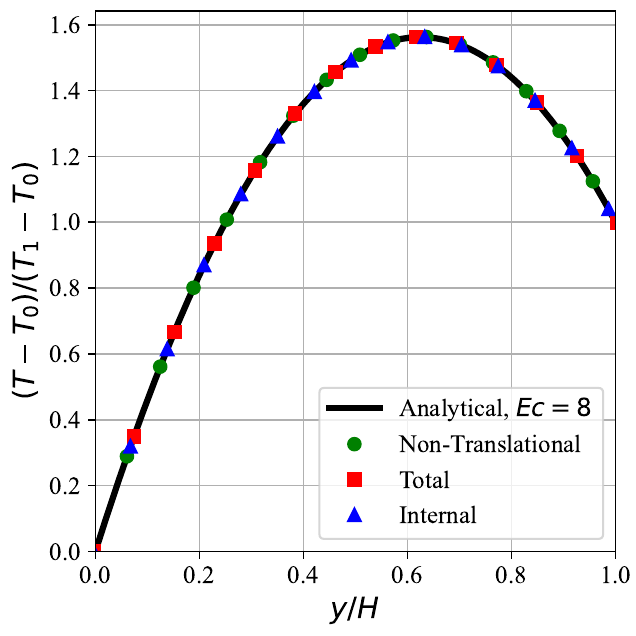}
    \caption{Temperature ratio for thermal Couette flow at $Ma=1$, $\mathrm{Pr=1}$, and $Ec=8$ for different realisations (symbols), plotted against analytical solution(line). }
    \label{fig:ThermalCouette_realisations}
\end{figure}

\begin{figure}
    \centering
    \includegraphics[width=0.9\linewidth]{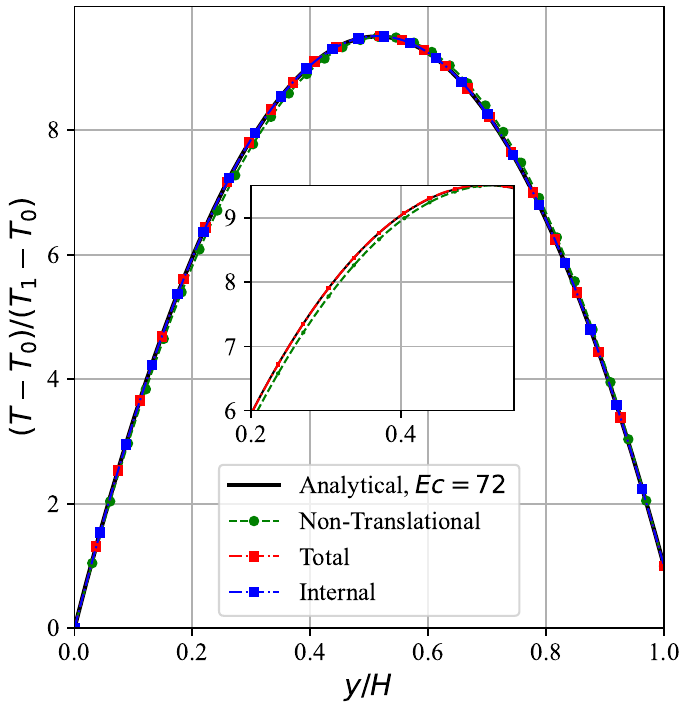}
    \caption{Temperature ratio for thermal Couette flow at $Ec=72$ for different realisations (symbols) on a D2Q16 lattice, plotted against analytical solution (line). Inset: Detail of plot showing error in temperature profile achieved using the non-translational realisation.}
    \label{fig:ThermalCouette_realisations_Ec72}
\end{figure}

\begin{figure}
    \centering
    \includegraphics[width=0.9\linewidth]{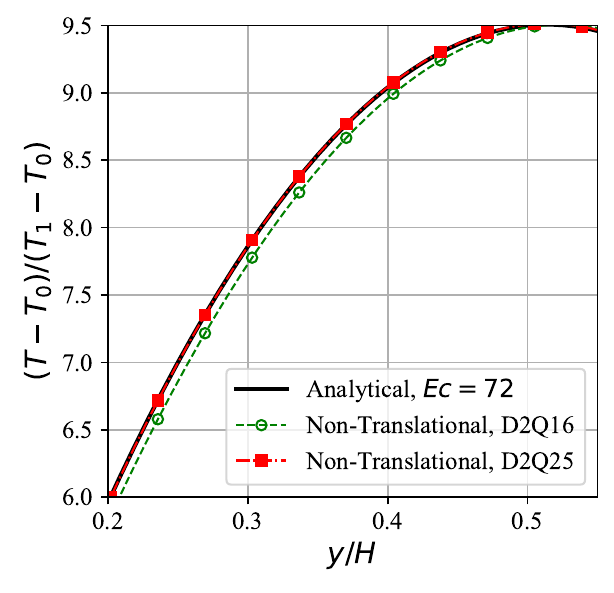}
    \caption{ Deviation in temperature profile resulting from the D2Q16 lattice visualised using high Eckert number thermal Couette flow. A shift to the D2Q25 lattice restores the fourth-order moments and leads to proper recovery of the full Navier Stokes Fourier system.}
    \label{fig:ThermalCouette_q16vq25_internal}
\end{figure}

To assess the proper recovery of the combined effect of viscous heating and thermal conductivity we next consider the thermal Couette flow. The configuration consists of a 2-D channel of height $L$ bounded by a stagnant wall (at the bottom) at temperature $T_0$ (set to $T_0=1$ here) and a moving wall (at the top), at velocity $u_0$ and temperature $T_1$ (set to $T_1=1.005$ here). 
The analytical solution for the temperature distribution in the channel is~\cite{liepmann1957},
\begin{equation}
    \frac{T - T_0}{T_1-T_0} = \frac{y}{L} + \frac{{\rm Pr} {\rm Ec}}{2}\frac{y}{L}\left(1-\frac{y}{L}\right),
\end{equation}
where the Eckert number is defined as ${\rm Ec} = u_0^2/(c_v+r)\Delta T$ 
with $\Delta T = T_1-T_0$. 
A first set of simulations was conducted at ${\rm Ec}=8$ using all splitting schemes, with (parameters). The results are displayed in Fig.~\ref{fig:ThermalCouette_realisations}.
Simulations with these set of parameters showed very good agreement with the analytical solution. To better show the effect of non-recovery of the fourth-order moment in the non-translational energy split a second set of simulation was conducted at ${\rm Ec}=72$ with stronger velocity and temperature gradients. The results are displayed in Fig.~\ref{fig:ThermalCouette_realisations_Ec72}.
It is observed that the D2Q16 lattice combined with the non-translational split is the only scheme that leads to visible deviations from the reference analytical solution. This is to be expected as, per discussions in the previous sections, the fourth-order moment of the first-distribution function is not recovered on this lattice, and therefore leads to errors in the diffusive heat flux. Note, however that due to the locally-adaptive nature of PonD, these deviations are considerably reduced as compared to static reference frame realizations like the LBM. Finally, to further prove the source of these deviations, a simulation was also conducted using a higher-order lattice, i.e. D2Q25, and using the same grid-size. Results are displayed in Fig.~\ref{fig:ThermalCouette_q16vq25_internal}, and we can see that restoring the fourth-order moment of the first-distribution function corrects these deviations.
\subsubsection{2-D test-case: High Mach astrophysical jet}
\begin{figure}
    \centering
    \includegraphics[width=1\linewidth]{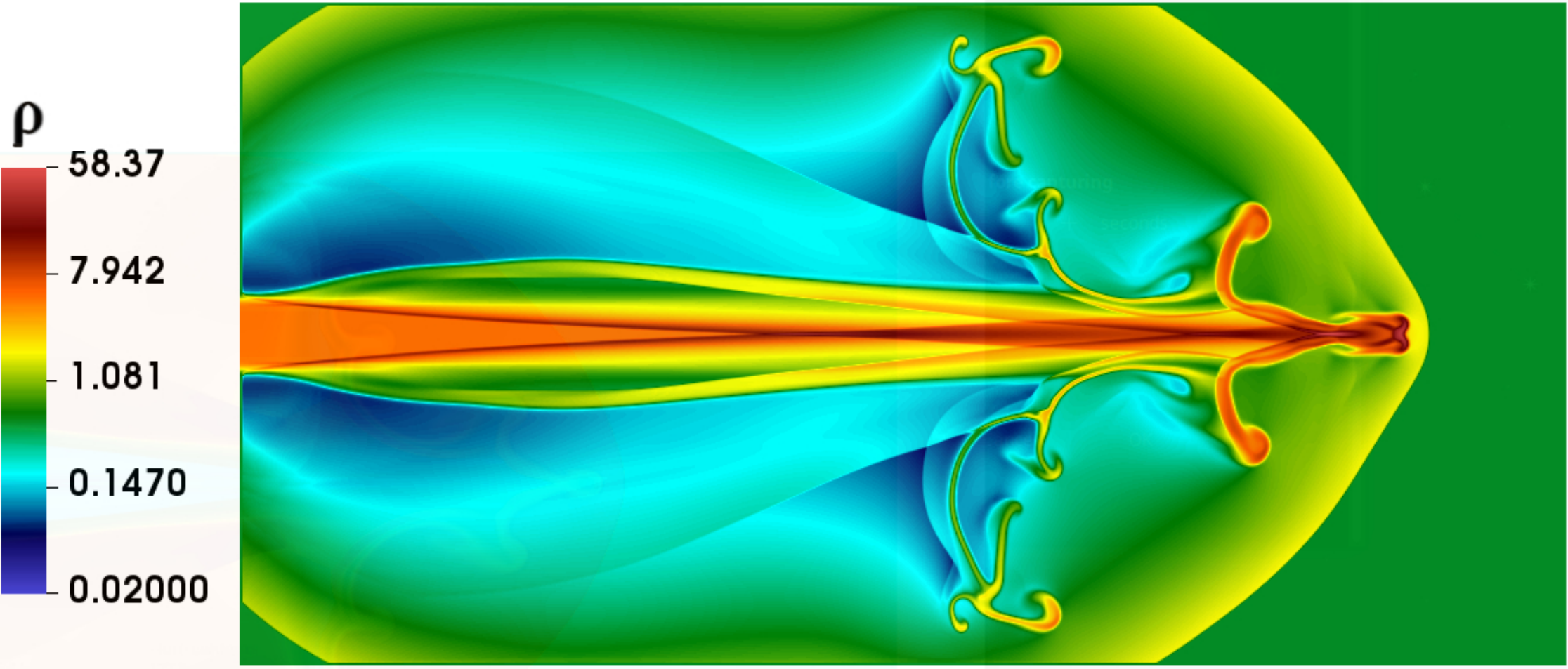}
    \includegraphics[width=1\linewidth]{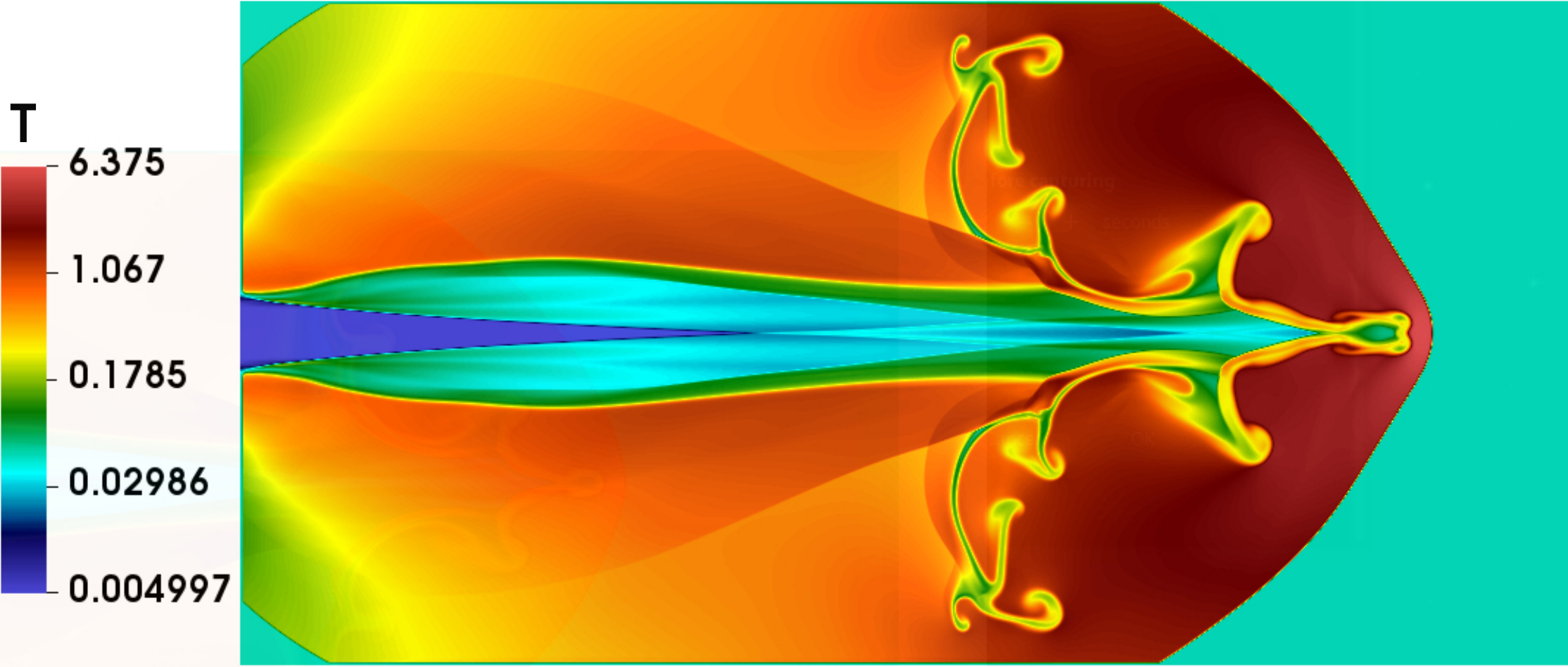}
    \includegraphics[width=1\linewidth]{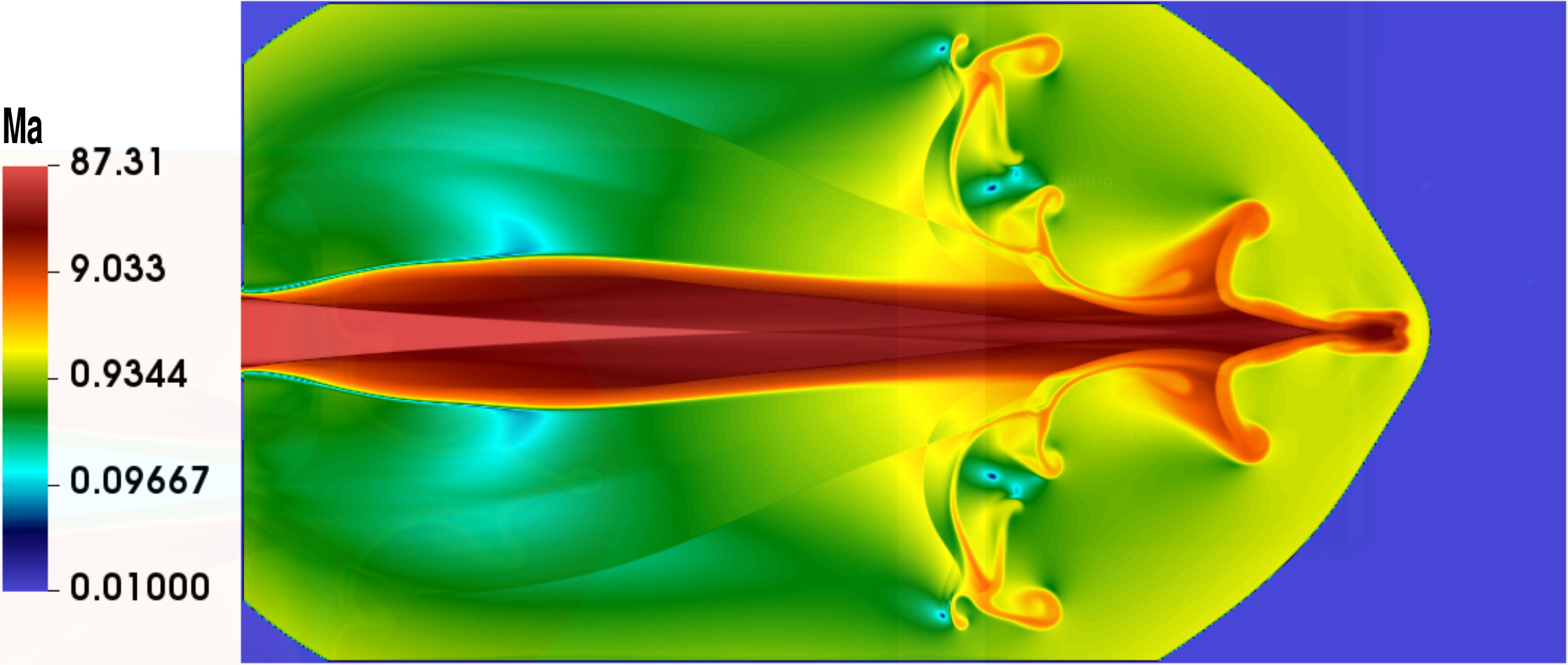}
    \caption{Astrophysical jet at Mach $80$. (Top) Density, (middle) temperature, and (bottom) Mach number.}
    \label{fig:astroJet}
\end{figure}

As a demonstration of applicability to high Mach numbers, we consider the case of an astrophysical jet of Mach $80$, without radiative cooling [13]. This case is an example of actual gas flows revealed from images of the Hubble Space Telescope and therefore is of high scientific interest. Furthermore, given the rather large variation in temperature and pressure it is a challenging configuration to run using any numerical scheme. Following the configuration in (), the jet is initialized as a semi-circular region on the left boundary : $$\{\rm{Ma}, \rho, T\}  = \{80, 5, 0.005\}. $$ 
The rest of the computational domain is at rest with $$ \{\rm{Ma}, \rho, T\}  = \{0, 50, 0.0005\}.$$ Outflow boundary conditions are used all around except on the left boundary where the prescribed conditions are imposed. The simulation was conducted with a resolution of $2000\times1000$ with the diameter of the initial jet, $D_{jet}=100.$  The simulation was conducted using the total energy split and D2Q16 only, as based on previous discussions it was selected as the more efficient and accurate realization. Figure \ref{fig:astroJet} shows the density, pressure and temperature field at lattice $t=240$.
\section{Conclusions and discussion}
With the advent of more reliable discrete kinetic schemes for compressible flow simulations, development and tuning of efficient and accurate realizations has become a central topic in the literature. In the present contribution we aimed at discussing the different possible realizations in the context of double distribution function models with a focus on high speed flows. To that end multi-scale analyses of all schemes, considering internal degrees of freedom, were conducted. All energy splits were shown to recover consistent hydrodynamics in agreement with the single distribution function. Furthermore it was shown that one of the energy splits, i.e. non-translational internal, required higher-order quadratures to properly satisfy all equilibrium moments needed for the Navier-Stokes-Fourier equations. Simulations of the thermal Couette flow confirmed the issues with the fourth-order equilibrium moments, as for higher Eckert number the D2Q16 lattice led to non-negligible deviations which were not present for the D2Q25 lattice. The internal energy split on the other hand, while recovering proper hydrodynamics with fourth-order quadratures, brings in source terms that are non-local in both space and time. This impacts both computational load, memory requirements and also numerical properties. In our simulations, the added cost of the higher-order quadrature was approximately $\sim 60\%$ as compared to a {fourth-order quadrature} --note that this number will go up in 3-D simulations-- while for the internal energy realization the source terms contributed to a $\sim 15\%$ increase in computation time as compared to the to total energy realization.


\begin{acknowledgments}
This work was supported by European Research Council (ERC) Advanced Grant  834763-PonD. 
Computational resources at the Swiss National  Super  Computing  Center  CSCS  were  provided  under the grant s1286.
\end{acknowledgments}

\bibliography{references}

\begin{thebibliography}{31}%
\makeatletter
\providecommand \@ifxundefined [1]{%
 \@ifx{#1\undefined}
}%
\providecommand \@ifnum [1]{%
 \ifnum #1\expandafter \@firstoftwo
 \else \expandafter \@secondoftwo
 \fi
}%
\providecommand \@ifx [1]{%
 \ifx #1\expandafter \@firstoftwo
 \else \expandafter \@secondoftwo
 \fi
}%
\providecommand \natexlab [1]{#1}%
\providecommand \enquote  [1]{``#1''}%
\providecommand \bibnamefont  [1]{#1}%
\providecommand \bibfnamefont [1]{#1}%
\providecommand \citenamefont [1]{#1}%
\providecommand \href@noop [0]{\@secondoftwo}%
\providecommand \href [0]{\begingroup \@sanitize@url \@href}%
\providecommand \@href[1]{\@@startlink{#1}\@@href}%
\providecommand \@@href[1]{\endgroup#1\@@endlink}%
\providecommand \@sanitize@url [0]{\catcode `\\12\catcode `\$12\catcode `\&12\catcode `\#12\catcode `\^12\catcode `\_12\catcode `\%12\relax}%
\providecommand \@@startlink[1]{}%
\providecommand \@@endlink[0]{}%
\providecommand \url  [0]{\begingroup\@sanitize@url \@url }%
\providecommand \@url [1]{\endgroup\@href {#1}{\urlprefix }}%
\providecommand \urlprefix  [0]{URL }%
\providecommand \Eprint [0]{\href }%
\providecommand \doibase [0]{https://doi.org/}%
\providecommand \selectlanguage [0]{\@gobble}%
\providecommand \bibinfo  [0]{\@secondoftwo}%
\providecommand \bibfield  [0]{\@secondoftwo}%
\providecommand \translation [1]{[#1]}%
\providecommand \BibitemOpen [0]{}%
\providecommand \bibitemStop [0]{}%
\providecommand \bibitemNoStop [0]{.\EOS\space}%
\providecommand \EOS [0]{\spacefactor3000\relax}%
\providecommand \BibitemShut  [1]{\csname bibitem#1\endcsname}%
\let\auto@bib@innerbib\@empty
\bibitem [{\citenamefont {Hosseini}\ \emph {et~al.}(2024)\citenamefont {Hosseini}, \citenamefont {Boivin}, \citenamefont {Th{\'e}venin},\ and\ \citenamefont {Karlin}}]{hosseini2024lattice}%
  \BibitemOpen
  \bibfield  {author} {\bibinfo {author} {\bibfnamefont {S.~A.}\ \bibnamefont {Hosseini}}, \bibinfo {author} {\bibfnamefont {P.}~\bibnamefont {Boivin}}, \bibinfo {author} {\bibfnamefont {D.}~\bibnamefont {Th{\'e}venin}},\ and\ \bibinfo {author} {\bibfnamefont {I.}~\bibnamefont {Karlin}},\ }\bibfield  {title} {\bibinfo {title} {Lattice boltzmann methods for combustion applications},\ }\href@noop {} {\bibfield  {journal} {\bibinfo  {journal} {Progress in Energy and Combustion Science}\ }\textbf {\bibinfo {volume} {102}},\ \bibinfo {pages} {101140} (\bibinfo {year} {2024})}\BibitemShut {NoStop}%
\bibitem [{\citenamefont {Philippi}\ \emph {et~al.}(2006)\citenamefont {Philippi}, \citenamefont {Hegele~Jr}, \citenamefont {Dos~Santos},\ and\ \citenamefont {Surmas}}]{philippi2006continuous}%
  \BibitemOpen
  \bibfield  {author} {\bibinfo {author} {\bibfnamefont {P.~C.}\ \bibnamefont {Philippi}}, \bibinfo {author} {\bibfnamefont {L.~A.}\ \bibnamefont {Hegele~Jr}}, \bibinfo {author} {\bibfnamefont {L.~O.}\ \bibnamefont {Dos~Santos}},\ and\ \bibinfo {author} {\bibfnamefont {R.}~\bibnamefont {Surmas}},\ }\bibfield  {title} {\bibinfo {title} {From the continuous to the lattice boltzmann equation: The discretization problem and thermal models},\ }\href@noop {} {\bibfield  {journal} {\bibinfo  {journal} {Physical Review E}\ }\textbf {\bibinfo {volume} {73}},\ \bibinfo {pages} {056702} (\bibinfo {year} {2006})}\BibitemShut {NoStop}%
\bibitem [{\citenamefont {Shan}\ \emph {et~al.}(2006)\citenamefont {Shan}, \citenamefont {Yuan},\ and\ \citenamefont {Chen}}]{shan2006kinetic}%
  \BibitemOpen
  \bibfield  {author} {\bibinfo {author} {\bibfnamefont {X.}~\bibnamefont {Shan}}, \bibinfo {author} {\bibfnamefont {X.-F.}\ \bibnamefont {Yuan}},\ and\ \bibinfo {author} {\bibfnamefont {H.}~\bibnamefont {Chen}},\ }\bibfield  {title} {\bibinfo {title} {Kinetic theory representation of hydrodynamics: a way beyond the navier--stokes equation},\ }\href@noop {} {\bibfield  {journal} {\bibinfo  {journal} {Journal of Fluid Mechanics}\ }\textbf {\bibinfo {volume} {550}},\ \bibinfo {pages} {413} (\bibinfo {year} {2006})}\BibitemShut {NoStop}%
\bibitem [{\citenamefont {Siebert}\ \emph {et~al.}(2008)\citenamefont {Siebert}, \citenamefont {Hegele~Jr},\ and\ \citenamefont {Philippi}}]{siebert2008lattice}%
  \BibitemOpen
  \bibfield  {author} {\bibinfo {author} {\bibfnamefont {D.}~\bibnamefont {Siebert}}, \bibinfo {author} {\bibfnamefont {L.}~\bibnamefont {Hegele~Jr}},\ and\ \bibinfo {author} {\bibfnamefont {P.~C.}\ \bibnamefont {Philippi}},\ }\bibfield  {title} {\bibinfo {title} {Lattice boltzmann equation linear stability analysis: Thermal and athermal models},\ }\href@noop {} {\bibfield  {journal} {\bibinfo  {journal} {Physical Review E}\ }\textbf {\bibinfo {volume} {77}},\ \bibinfo {pages} {026707} (\bibinfo {year} {2008})}\BibitemShut {NoStop}%
\bibitem [{\citenamefont {Frapolli}\ \emph {et~al.}(2014)\citenamefont {Frapolli}, \citenamefont {Chikatamarla},\ and\ \citenamefont {Karlin}}]{frapolli2014multispeed}%
  \BibitemOpen
  \bibfield  {author} {\bibinfo {author} {\bibfnamefont {N.}~\bibnamefont {Frapolli}}, \bibinfo {author} {\bibfnamefont {S.}~\bibnamefont {Chikatamarla}},\ and\ \bibinfo {author} {\bibfnamefont {I.}~\bibnamefont {Karlin}},\ }\bibfield  {title} {\bibinfo {title} {Multispeed entropic lattice boltzmann model for thermal flows},\ }\href@noop {} {\bibfield  {journal} {\bibinfo  {journal} {Physical Review E}\ }\textbf {\bibinfo {volume} {90}},\ \bibinfo {pages} {043306} (\bibinfo {year} {2014})}\BibitemShut {NoStop}%
\bibitem [{\citenamefont {He}\ \emph {et~al.}(1998)\citenamefont {He}, \citenamefont {Chen},\ and\ \citenamefont {Doolen}}]{he1998novel}%
  \BibitemOpen
  \bibfield  {author} {\bibinfo {author} {\bibfnamefont {X.}~\bibnamefont {He}}, \bibinfo {author} {\bibfnamefont {S.}~\bibnamefont {Chen}},\ and\ \bibinfo {author} {\bibfnamefont {G.~D.}\ \bibnamefont {Doolen}},\ }\bibfield  {title} {\bibinfo {title} {A novel thermal model for the lattice boltzmann method in incompressible limit},\ }\href@noop {} {\bibfield  {journal} {\bibinfo  {journal} {Journal of computational physics}\ }\textbf {\bibinfo {volume} {146}},\ \bibinfo {pages} {282} (\bibinfo {year} {1998})}\BibitemShut {NoStop}%
\bibitem [{\citenamefont {Guo}\ \emph {et~al.}(2007)\citenamefont {Guo}, \citenamefont {Zheng}, \citenamefont {Shi},\ and\ \citenamefont {Zhao}}]{guo2007thermal}%
  \BibitemOpen
  \bibfield  {author} {\bibinfo {author} {\bibfnamefont {Z.}~\bibnamefont {Guo}}, \bibinfo {author} {\bibfnamefont {C.}~\bibnamefont {Zheng}}, \bibinfo {author} {\bibfnamefont {B.}~\bibnamefont {Shi}},\ and\ \bibinfo {author} {\bibfnamefont {T.}~\bibnamefont {Zhao}},\ }\bibfield  {title} {\bibinfo {title} {Thermal lattice boltzmann equation for low mach number flows: Decoupling model},\ }\href@noop {} {\bibfield  {journal} {\bibinfo  {journal} {Physical Review E}\ }\textbf {\bibinfo {volume} {75}},\ \bibinfo {pages} {036704} (\bibinfo {year} {2007})}\BibitemShut {NoStop}%
\bibitem [{\citenamefont {Karlin}\ \emph {et~al.}(2013)\citenamefont {Karlin}, \citenamefont {Sichau},\ and\ \citenamefont {Chikatamarla}}]{karlin2013consistent}%
  \BibitemOpen
  \bibfield  {author} {\bibinfo {author} {\bibfnamefont {I.}~\bibnamefont {Karlin}}, \bibinfo {author} {\bibfnamefont {D.}~\bibnamefont {Sichau}},\ and\ \bibinfo {author} {\bibfnamefont {S.}~\bibnamefont {Chikatamarla}},\ }\bibfield  {title} {\bibinfo {title} {Consistent two-population lattice boltzmann model for thermal flows},\ }\href@noop {} {\bibfield  {journal} {\bibinfo  {journal} {Physical Review E}\ }\textbf {\bibinfo {volume} {88}},\ \bibinfo {pages} {063310} (\bibinfo {year} {2013})}\BibitemShut {NoStop}%
\bibitem [{\citenamefont {Asinari}\ and\ \citenamefont {Karlin}(2010)}]{asinari2010quasiequilibrium}%
  \BibitemOpen
  \bibfield  {author} {\bibinfo {author} {\bibfnamefont {P.}~\bibnamefont {Asinari}}\ and\ \bibinfo {author} {\bibfnamefont {I.~V.}\ \bibnamefont {Karlin}},\ }\bibfield  {title} {\bibinfo {title} {Quasiequilibrium lattice boltzmann models with tunable bulk viscosity for enhancing stability},\ }\href@noop {} {\bibfield  {journal} {\bibinfo  {journal} {Physical Review E}\ }\textbf {\bibinfo {volume} {81}},\ \bibinfo {pages} {016702} (\bibinfo {year} {2010})}\BibitemShut {NoStop}%
\bibitem [{\citenamefont {Hosseini}\ and\ \citenamefont {Karlin}(2023)}]{hosseini2023lattice}%
  \BibitemOpen
  \bibfield  {author} {\bibinfo {author} {\bibfnamefont {S.}~\bibnamefont {Hosseini}}\ and\ \bibinfo {author} {\bibfnamefont {I.}~\bibnamefont {Karlin}},\ }\bibfield  {title} {\bibinfo {title} {Lattice boltzmann for non-ideal fluids: Fundamentals and practice},\ }\href@noop {} {\bibfield  {journal} {\bibinfo  {journal} {Physics Reports}\ }\textbf {\bibinfo {volume} {1030}},\ \bibinfo {pages} {1} (\bibinfo {year} {2023})}\BibitemShut {NoStop}%
\bibitem [{\citenamefont {Rykov}(1975)}]{rykov1975model}%
  \BibitemOpen
  \bibfield  {author} {\bibinfo {author} {\bibfnamefont {V.}~\bibnamefont {Rykov}},\ }\bibfield  {title} {\bibinfo {title} {A model kinetic equation for a gas with rotational degrees of freedom},\ }\href@noop {} {\bibfield  {journal} {\bibinfo  {journal} {Fluid Dynamics}\ }\textbf {\bibinfo {volume} {10}},\ \bibinfo {pages} {959} (\bibinfo {year} {1975})}\BibitemShut {NoStop}%
\bibitem [{\citenamefont {Kolluru}\ \emph {et~al.}(2020)\citenamefont {Kolluru}, \citenamefont {Atif},\ and\ \citenamefont {Ansumali}}]{kolluru2020extended}%
  \BibitemOpen
  \bibfield  {author} {\bibinfo {author} {\bibfnamefont {P.~K.}\ \bibnamefont {Kolluru}}, \bibinfo {author} {\bibfnamefont {M.}~\bibnamefont {Atif}},\ and\ \bibinfo {author} {\bibfnamefont {S.}~\bibnamefont {Ansumali}},\ }\bibfield  {title} {\bibinfo {title} {Extended bgk model for diatomic gases},\ }\href@noop {} {\bibfield  {journal} {\bibinfo  {journal} {Journal of Computational Science}\ }\textbf {\bibinfo {volume} {45}},\ \bibinfo {pages} {101179} (\bibinfo {year} {2020})}\BibitemShut {NoStop}%
\bibitem [{\citenamefont {Gorban}\ and\ \citenamefont {Karlin}(1994)}]{gorban1994general}%
  \BibitemOpen
  \bibfield  {author} {\bibinfo {author} {\bibfnamefont {A.~N.}\ \bibnamefont {Gorban}}\ and\ \bibinfo {author} {\bibfnamefont {I.~V.}\ \bibnamefont {Karlin}},\ }\bibfield  {title} {\bibinfo {title} {General approach to constructing models of the boltzmann equation},\ }\href@noop {} {\bibfield  {journal} {\bibinfo  {journal} {Physica A: Statistical Mechanics and its Applications}\ }\textbf {\bibinfo {volume} {206}},\ \bibinfo {pages} {401} (\bibinfo {year} {1994})}\BibitemShut {NoStop}%
\bibitem [{\citenamefont {Ansumali}\ \emph {et~al.}(2007)\citenamefont {Ansumali}, \citenamefont {Arcidiacono}, \citenamefont {Chikatamarla}, \citenamefont {Prasianakis}, \citenamefont {Gorban},\ and\ \citenamefont {Karlin}}]{ansumali2007quasi}%
  \BibitemOpen
  \bibfield  {author} {\bibinfo {author} {\bibfnamefont {S.}~\bibnamefont {Ansumali}}, \bibinfo {author} {\bibfnamefont {S.}~\bibnamefont {Arcidiacono}}, \bibinfo {author} {\bibfnamefont {S.}~\bibnamefont {Chikatamarla}}, \bibinfo {author} {\bibfnamefont {N.}~\bibnamefont {Prasianakis}}, \bibinfo {author} {\bibfnamefont {A.}~\bibnamefont {Gorban}},\ and\ \bibinfo {author} {\bibfnamefont {I.}~\bibnamefont {Karlin}},\ }\bibfield  {title} {\bibinfo {title} {Quasi-equilibrium lattice boltzmann method},\ }\href@noop {} {\bibfield  {journal} {\bibinfo  {journal} {The European Physical Journal B}\ }\textbf {\bibinfo {volume} {56}},\ \bibinfo {pages} {135} (\bibinfo {year} {2007})}\BibitemShut {NoStop}%
\bibitem [{\citenamefont {Dorschner}\ \emph {et~al.}(2018)\citenamefont {Dorschner}, \citenamefont {B{\"o}sch},\ and\ \citenamefont {Karlin}}]{dorschner2018particles}%
  \BibitemOpen
  \bibfield  {author} {\bibinfo {author} {\bibfnamefont {B.}~\bibnamefont {Dorschner}}, \bibinfo {author} {\bibfnamefont {F.}~\bibnamefont {B{\"o}sch}},\ and\ \bibinfo {author} {\bibfnamefont {I.~V.}\ \bibnamefont {Karlin}},\ }\bibfield  {title} {\bibinfo {title} {Particles on demand for kinetic theory},\ }\href@noop {} {\bibfield  {journal} {\bibinfo  {journal} {Physical review letters}\ }\textbf {\bibinfo {volume} {121}},\ \bibinfo {pages} {130602} (\bibinfo {year} {2018})}\BibitemShut {NoStop}%
\bibitem [{\citenamefont {Bhatnagar}\ \emph {et~al.}(1954)\citenamefont {Bhatnagar}, \citenamefont {Gross},\ and\ \citenamefont {Krook}}]{bhatnagar1954model}%
  \BibitemOpen
  \bibfield  {author} {\bibinfo {author} {\bibfnamefont {P.~L.}\ \bibnamefont {Bhatnagar}}, \bibinfo {author} {\bibfnamefont {E.~P.}\ \bibnamefont {Gross}},\ and\ \bibinfo {author} {\bibfnamefont {M.}~\bibnamefont {Krook}},\ }\bibfield  {title} {\bibinfo {title} {A model for collision processes in gases. i. small amplitude processes in charged and neutral one-component systems},\ }\href@noop {} {\bibfield  {journal} {\bibinfo  {journal} {Physical review}\ }\textbf {\bibinfo {volume} {94}},\ \bibinfo {pages} {511} (\bibinfo {year} {1954})}\BibitemShut {NoStop}%
\bibitem [{\citenamefont {Li}\ \emph {et~al.}(2007)\citenamefont {Li}, \citenamefont {He}, \citenamefont {Wang},\ and\ \citenamefont {Tao}}]{li2007coupled}%
  \BibitemOpen
  \bibfield  {author} {\bibinfo {author} {\bibfnamefont {Q.}~\bibnamefont {Li}}, \bibinfo {author} {\bibfnamefont {Y.}~\bibnamefont {He}}, \bibinfo {author} {\bibfnamefont {Y.}~\bibnamefont {Wang}},\ and\ \bibinfo {author} {\bibfnamefont {W.}~\bibnamefont {Tao}},\ }\bibfield  {title} {\bibinfo {title} {Coupled double-distribution-function lattice boltzmann method for the compressible navier-stokes equations},\ }\href@noop {} {\bibfield  {journal} {\bibinfo  {journal} {Physical Review E}\ }\textbf {\bibinfo {volume} {76}},\ \bibinfo {pages} {056705} (\bibinfo {year} {2007})}\BibitemShut {NoStop}%
\bibitem [{\citenamefont {Prasianakis}\ and\ \citenamefont {Karlin}(2007)}]{prasianakis2007lattice}%
  \BibitemOpen
  \bibfield  {author} {\bibinfo {author} {\bibfnamefont {N.~I.}\ \bibnamefont {Prasianakis}}\ and\ \bibinfo {author} {\bibfnamefont {I.~V.}\ \bibnamefont {Karlin}},\ }\bibfield  {title} {\bibinfo {title} {Lattice boltzmann method for thermal flow simulation on standard lattices},\ }\href@noop {} {\bibfield  {journal} {\bibinfo  {journal} {Physical Review E}\ }\textbf {\bibinfo {volume} {76}},\ \bibinfo {pages} {016702} (\bibinfo {year} {2007})}\BibitemShut {NoStop}%
\bibitem [{\citenamefont {Hosseini}\ \emph {et~al.}(2020)\citenamefont {Hosseini}, \citenamefont {Darabiha},\ and\ \citenamefont {Th{\'{e}}venin}}]{hosseini2020compressibility}%
  \BibitemOpen
  \bibfield  {author} {\bibinfo {author} {\bibfnamefont {S.~A.}\ \bibnamefont {Hosseini}}, \bibinfo {author} {\bibfnamefont {N.}~\bibnamefont {Darabiha}},\ and\ \bibinfo {author} {\bibfnamefont {D.}~\bibnamefont {Th{\'{e}}venin}},\ }\bibfield  {title} {\bibinfo {title} {Compressibility in lattice {B}oltzmann on standard stencils: effects of deviation from reference temperature},\ }\href {https://doi.org/10.1098/rsta.2019.0399} {\bibfield  {journal} {\bibinfo  {journal} {Philosophical Transactions of the Royal Society A: Mathematical, Physical and Engineering Sciences}\ }\textbf {\bibinfo {volume} {378}},\ \bibinfo {pages} {20190399} (\bibinfo {year} {2020})}\BibitemShut {NoStop}%
\bibitem [{\citenamefont {Frapolli}\ \emph {et~al.}(2015)\citenamefont {Frapolli}, \citenamefont {Chikatamarla},\ and\ \citenamefont {Karlin}}]{frapolli2015}%
  \BibitemOpen
  \bibfield  {author} {\bibinfo {author} {\bibfnamefont {N.}~\bibnamefont {Frapolli}}, \bibinfo {author} {\bibfnamefont {S.~S.}\ \bibnamefont {Chikatamarla}},\ and\ \bibinfo {author} {\bibfnamefont {I.~V.}\ \bibnamefont {Karlin}},\ }\bibfield  {title} {\bibinfo {title} {Entropic lattice boltzmann model for compressible flows},\ }\href@noop {} {\bibfield  {journal} {\bibinfo  {journal} {Physical Review E}\ }\textbf {\bibinfo {volume} {92}},\ \bibinfo {pages} {061301} (\bibinfo {year} {2015})}\BibitemShut {NoStop}%
\bibitem [{\citenamefont {Kauf}(2011)}]{kauf2011multi}%
  \BibitemOpen
  \bibfield  {author} {\bibinfo {author} {\bibfnamefont {P.}~\bibnamefont {Kauf}},\ }\href@noop {} {\emph {\bibinfo {title} {Multi-scale approximation models for the Boltzmann equation}}}\ (\bibinfo  {publisher} {{ETH} Zurich},\ \bibinfo {year} {2011})\BibitemShut {NoStop}%
\bibitem [{\citenamefont {K{\"o}llermeier}(2013)}]{kollermeier2013hyperbolic}%
  \BibitemOpen
  \bibfield  {author} {\bibinfo {author} {\bibfnamefont {J.}~\bibnamefont {K{\"o}llermeier}},\ }\bibfield  {title} {\bibinfo {title} {Hyperbolic approximation of kinetic equations using quadrature-based projection methods},\ }\href@noop {} {\bibfield  {journal} {\bibinfo  {journal} {Master's thesis, {RWTH} Aachen University}\ } (\bibinfo {year} {2013})}\BibitemShut {NoStop}%
\bibitem [{\citenamefont {Grad}(1949)}]{grad_kinetic_1949}%
  \BibitemOpen
  \bibfield  {author} {\bibinfo {author} {\bibfnamefont {H.}~\bibnamefont {Grad}},\ }\bibfield  {title} {\bibinfo {title} {On the kinetic theory of rarefied gases},\ }\href@noop {} {\bibfield  {journal} {\bibinfo  {journal} {Communications on pure and applied mathematics}\ }\textbf {\bibinfo {volume} {2}},\ \bibinfo {pages} {331} (\bibinfo {year} {1949})}\BibitemShut {NoStop}%
\bibitem [{\citenamefont {Zipunova}\ \emph {et~al.}(2021)\citenamefont {Zipunova}, \citenamefont {Perepelkina}, \citenamefont {Zakirov},\ and\ \citenamefont {Khilkov}}]{Zipunova_Reg2021}%
  \BibitemOpen
  \bibfield  {author} {\bibinfo {author} {\bibfnamefont {E.}~\bibnamefont {Zipunova}}, \bibinfo {author} {\bibfnamefont {A.}~\bibnamefont {Perepelkina}}, \bibinfo {author} {\bibfnamefont {A.}~\bibnamefont {Zakirov}},\ and\ \bibinfo {author} {\bibfnamefont {S.}~\bibnamefont {Khilkov}},\ }\bibfield  {title} {\bibinfo {title} {Regularization and the particles-on-demand method for the solution of the discrete {B}oltzmann equation},\ }\href {https://doi.org/https://doi.org/10.1016/j.jocs.2021.101376} {\bibfield  {journal} {\bibinfo  {journal} {Journal of Computational Science}\ }\textbf {\bibinfo {volume} {53}},\ \bibinfo {pages} {101376} (\bibinfo {year} {2021})}\BibitemShut {NoStop}%
\bibitem [{\citenamefont {Kallikounis}\ \emph {et~al.}(2022)\citenamefont {Kallikounis}, \citenamefont {Dorschner},\ and\ \citenamefont {Karlin}}]{kallikounis2022particles}%
  \BibitemOpen
  \bibfield  {author} {\bibinfo {author} {\bibfnamefont {N.}~\bibnamefont {Kallikounis}}, \bibinfo {author} {\bibfnamefont {B.}~\bibnamefont {Dorschner}},\ and\ \bibinfo {author} {\bibfnamefont {I.~V.}\ \bibnamefont {Karlin}},\ }\bibfield  {title} {\bibinfo {title} {Particles on demand for flows with strong discontinuities},\ }\href@noop {} {\bibfield  {journal} {\bibinfo  {journal} {Physical Review E}\ }\textbf {\bibinfo {volume} {16}},\ \bibinfo {pages} {015301} (\bibinfo {year} {2022})}\BibitemShut {NoStop}%
\bibitem [{\citenamefont {Sawant}\ \emph {et~al.}(2022)\citenamefont {Sawant}, \citenamefont {Dorschner},\ and\ \citenamefont {Karlin}}]{sawantPonD_2022}%
  \BibitemOpen
  \bibfield  {author} {\bibinfo {author} {\bibfnamefont {N.}~\bibnamefont {Sawant}}, \bibinfo {author} {\bibfnamefont {B.}~\bibnamefont {Dorschner}},\ and\ \bibinfo {author} {\bibfnamefont {I.~V.}\ \bibnamefont {Karlin}},\ }\bibfield  {title} {\bibinfo {title} {Detonation modeling with the particles on demand method},\ }\href {https://doi.org/10.1063/5.0095122} {\bibfield  {journal} {\bibinfo  {journal} {AIP Advances}\ }\textbf {\bibinfo {volume} {12}},\ \bibinfo {pages} {075107} (\bibinfo {year} {2022})}\BibitemShut {NoStop}%
\bibitem [{\citenamefont {Ji}\ \emph {et~al.}(2024)\citenamefont {Ji}, \citenamefont {Hosseini}, \citenamefont {Dorschner}, \citenamefont {Luo},\ and\ \citenamefont {Karlin}}]{ji2024eulerian}%
  \BibitemOpen
  \bibfield  {author} {\bibinfo {author} {\bibfnamefont {Y.}~\bibnamefont {Ji}}, \bibinfo {author} {\bibfnamefont {S.~A.}\ \bibnamefont {Hosseini}}, \bibinfo {author} {\bibfnamefont {B.}~\bibnamefont {Dorschner}}, \bibinfo {author} {\bibfnamefont {K.~H.}\ \bibnamefont {Luo}},\ and\ \bibinfo {author} {\bibfnamefont {I.~V.}\ \bibnamefont {Karlin}},\ }\bibfield  {title} {\bibinfo {title} {Eulerian discrete kinetic framework in comoving reference frame for hypersonic flows},\ }\href@noop {} {\bibfield  {journal} {\bibinfo  {journal} {Journal of Fluid Mechanics}\ }\textbf {\bibinfo {volume} {983}},\ \bibinfo {pages} {A11} (\bibinfo {year} {2024})}\BibitemShut {NoStop}%
\bibitem [{\citenamefont {Dellar}(2001)}]{dellar2001bulk}%
  \BibitemOpen
  \bibfield  {author} {\bibinfo {author} {\bibfnamefont {P.~J.}\ \bibnamefont {Dellar}},\ }\bibfield  {title} {\bibinfo {title} {Bulk and shear viscosities in lattice boltzmann equations},\ }\href@noop {} {\bibfield  {journal} {\bibinfo  {journal} {Physical Review E}\ }\textbf {\bibinfo {volume} {64}},\ \bibinfo {pages} {031203} (\bibinfo {year} {2001})}\BibitemShut {NoStop}%
\bibitem [{\citenamefont {Buick}\ \emph {et~al.}(2000)\citenamefont {Buick}, \citenamefont {Buckley}, \citenamefont {Greated},\ and\ \citenamefont {Gilbert}}]{buick2000lattice}%
  \BibitemOpen
  \bibfield  {author} {\bibinfo {author} {\bibfnamefont {J.}~\bibnamefont {Buick}}, \bibinfo {author} {\bibfnamefont {C.}~\bibnamefont {Buckley}}, \bibinfo {author} {\bibfnamefont {C.}~\bibnamefont {Greated}},\ and\ \bibinfo {author} {\bibfnamefont {J.}~\bibnamefont {Gilbert}},\ }\bibfield  {title} {\bibinfo {title} {Lattice {B}oltzmann {BGK} simulation of nonlinear sound waves: the development of a shock front},\ }\href@noop {} {\bibfield  {journal} {\bibinfo  {journal} {Journal of {P}hysics {A}: {M}athematical and General}\ }\textbf {\bibinfo {volume} {33}},\ \bibinfo {pages} {3917} (\bibinfo {year} {2000})}\BibitemShut {NoStop}%
\bibitem [{\citenamefont {Lamb}(1924)}]{lamb1924hydrodynamics}%
  \BibitemOpen
  \bibfield  {author} {\bibinfo {author} {\bibfnamefont {H.}~\bibnamefont {Lamb}},\ }\href@noop {} {\emph {\bibinfo {title} {Hydrodynamics}}}\ (\bibinfo  {publisher} {University Press},\ \bibinfo {year} {1924})\BibitemShut {NoStop}%
\bibitem [{\citenamefont {Liepmann}\ and\ \citenamefont {Roshko}(1957)}]{liepmann1957}%
  \BibitemOpen
  \bibfield  {author} {\bibinfo {author} {\bibfnamefont {H.~W.}\ \bibnamefont {Liepmann}}\ and\ \bibinfo {author} {\bibfnamefont {A.}~\bibnamefont {Roshko}},\ }\href@noop {} {\emph {\bibinfo {title} {Elements of gasdynamics}}}\ (\bibinfo  {publisher} {Wiley, New York, 1957},\ \bibinfo {year} {1957})\BibitemShut {NoStop}%
\end{thebibliography}%

\appendix
\section{Multi-scale analysis of double distribution function models}
\label{app:CE}
Let us consider the following general system of equations,
\begin{equation}
	\mathcal{D}_t \{f, g\} = \frac{1}{\tau}\left(\{f^{\rm eq}, g^{\rm eq}\} - \{f, g\}\right) + \{\mathcal{F}, \mathcal{G}\},
\end{equation}
where, $\mathcal{D}_t=\partial_t + \bm{v}\cdot\bm{\nabla}$, $\mathcal{F}=0$ for all splitting approaches, and $\mathcal{G}$ is only non-zero for the internal energy splitting with,
\begin{equation}
    \mathcal{G} = -f(\bm{v}-\bm{u})\cdot(\partial_t\bm{u} + \bm{v}\cdot\bm{\nabla}\bm{u}).
\end{equation}
For the multi-scale analysis let us introduce the following parameters: characteristic flow velocity $\mathcal{U}$, characteristic flow scale $\mathcal{L}$, characteristic flow time $\mathcal{T}=\mathcal{L}/\mathcal{U}$, characteristic density $\bar{\rho}$, speed of sound of ideal gas $c_s=\sqrt{\gamma r T}$. With these, the variables are reduced as follows (primes denote non-dimensional variables): time $t=\mathcal{T}t'$, space $\bm{x}=\mathcal{L}\bm{x}'$, flow velocity $\bm{u}=\mathcal{U}\bm{u}'$, particle velocity $\bm{v}=c_s\bm{v}'$, density $\rho$=$\bar{\rho}\rho'$, distribution function $f=\bar{\rho}c_s^{-3}f'$.
Furthermore, the following non-dimensional groups are introduced: Knudsen number ${\rm Kn}={\tau c_s}/{\mathcal{L}}$ and Mach number ${\rm Ma}={\mathcal{U}}/{c_s}$. With this, the equations are rescaled as follows:
\begin{equation}
	{\rm Kn}\,\mathcal{D}_t' \{f', g'\} = \frac{1}{\tau'}\left(\{f^{\rm eq'}, g^{\rm eq'}\} - \{f', g'\}\right) + \{\mathcal{F}', \mathcal{G}'\},
\end{equation}
Assuming ${\rm Kn}\sim\epsilon$ and dropping the primes for the sake of readability,
	\begin{equation}
	    \epsilon \mathcal{D}_t \{f, g\} = \frac{1}{\tau}\left(\{f^{\rm eq}, g^{\rm eq}\} - \{f, g\}\right) + \{\mathcal{F}, \mathcal{G}\}.
	\end{equation}
Then introducing multi-scale expansions:
	\begin{equation}
	    \{f, g\} = \{f^{(0)}, g^{(0)}\} + \epsilon \{f^{(1)}, g^{(1)}\} + \epsilon^2 \{f^{(2)}, g^{(2)}\} + O(\epsilon^3),
	\end{equation}
and,
	\begin{equation}
	    \{\mathcal{F}, \mathcal{G}\} = \epsilon \{\mathcal{F}^{(1)}, \mathcal{G}^{(1)}\} + \epsilon^2 \{\mathcal{F}^{(2)}, \mathcal{G}^{(2)}\} + O(\epsilon^3),
	\end{equation}
the following equations are recovered at scales $\epsilon$ and $\epsilon^2$:
	\begin{subequations}
	\begin{align}
	\epsilon &: \mathcal{D}_{t}^{(1)} \{f^{(0)}, g^{(0)}\} = -\frac{1}{\tau} \{f^{(1)}, g^{(1)}\} + \{\mathcal{F}^{(1)}, \mathcal{G}^{(1)}\}, \label{Eq:CE_Eq_orders_1}\\
	\epsilon^2 &: \partial_t^{(2)}\{f^{(0)}, g^{(0)}\} + \mathcal{D}_{t}^{(1)}\{f^{(1)}, g^{(1)}\} = -\frac{1}{\tau} \{f^{(2)}, g^{(2)}\} \nonumber \\  &+ \{\mathcal{F}^{(2)}, \mathcal{G}^{(2)}\}, \label{Eq:CE_Eq_orders_2}
	\end{align}
    \label{Eq:CE_Eq_orders}
    \end{subequations}
with $\{f^{(0)}, g^{(0)}\}=\{f^{\rm eq}, g^{\rm eq}\}$. Note that for this system, the solvability conditions are:
	\begin{subequations}
	\begin{align}
    \forall k>0, & \int_{\mathbb{R}^D} f^{(k)} d\bm{v}= 0,\\
	\forall k>0, & \int_{\mathbb{R}^D} \bm{v} f^{(k)} d\bm{v} = 0.
	\end{align}
    \end{subequations}
In addition, depending on the splitting, solvability conditions on energy are:
\begin{align}
	   \int_{\mathbb{R}^D} g^{(k)}d\bm{v} = 0,\, & \forall k>0,\\
	   \int_{\mathbb{R}^D} g^{(k)}d\bm{v} = 0,\, & \forall k>0,\\
    \int_{\mathbb{R}^D} \left(g^{(k)} + \frac{\bm{v}^2}{2} f^{(k)}\right) d\bm{v} = 0,\, & \forall k>0.
\end{align}
Taking the moments, $\int\{f, \bm{v} f\}d\bm{v}$, of the Chapman-Enskog-expanded equations at order $\epsilon$:
	\begin{eqnarray}
	    \partial_t^{(1)}\rho + \bm{\nabla}\cdot\rho \bm{u} &=& 0,\label{eq:approach2_continuity1}\\
	    \partial_t^{(1)}\rho \bm{u} + \bm{\nabla}\cdot\rho \bm{u}\otimes\bm{u} + \bm{\nabla}\cdot p\bm{I} &=& 0.\label{eq:approach2_NS1}
	\end{eqnarray}
For the energy balance equation, taking these moments $\int_{\mathbb{R}^D}\{g, g + \frac{\bm{u}^2}{2}f, g + \frac{\bm{v}^2}{2}f\}d\bm{v}$,
\begin{equation}
	\partial_t^{(1)}E + \bm{\nabla}\cdot E \bm{u} + \bm{\nabla}\cdot p\bm{u} = 0.
\end{equation}
Using the Euler level momentum balance and continuity equations, a balance equation for kinetic energy can be derived as,
\begin{equation}
    \partial_t^{(1)}K + \bm{\nabla}\cdot K \bm{u} + \bm{u}\cdot \bm{\nabla}p = 0,
\end{equation}
which in turn can be used to derive a balance equation for internal energy,
\begin{equation}
    \partial_t^{(1)} U + \bm{\nabla}\cdot U \bm{u} + p \bm{\nabla}\cdot\bm{u} = 0,
\end{equation}
and pressure,
\begin{equation}
    \partial_t^{(1)} p + \bm{\nabla}\cdot p \bm{u} + \frac{r}{c_v} p \bm{\nabla}\cdot\bm{u} = 0,\label{eq:pressure_euler_CE}
\end{equation}
where
\begin{equation}
        \frac{\partial U}{\partial T} = \rho c_v.
\end{equation}
Going up one order in $\epsilon$, at order $\epsilon^2$ the continuity equation is:
	\begin{equation}
	    \partial_t^{(2)}\rho = 0,\label{eq:approach2_continuity2}
	\end{equation}
while for the momentum balance equation one has:
    \begin{equation}
        \partial_t^{(2)}\rho \bm{u} + \bm{\nabla}\left(\int_{\mathbb{R}^D}\bm{v}\otimes\bm{v} f^{(1)}\right) = 0.
    \end{equation}
The second term can be further expanded using the first-order-in-$\epsilon$ equation as:
    \begin{multline}
        \int_{\mathbb{R}^D}\bm{v}\otimes\bm{v} f^{(1)}d\bm{v} =\\ -\tau\left[\partial_t^{(1)}\left(\rho\bm{u}\otimes\bm{u} + p\bm{I}\right) + \bm{\nabla}\cdot \rho\bm{u}\otimes\bm{u}\otimes\bm{u} \right. \\ \left. + \bm{\nabla}p\bm{u} + \bm{\nabla}p\bm{u}^{\dagger} + \bm{I}\bm{\nabla}\cdot p\bm{u}\right].
    \end{multline}
Next we expand the first term as:
    \begin{multline}
        \partial_t^{(1)}\left(\rho\bm{u}\otimes\bm{u} + p\bm{I}\right) = \bm{u}\otimes\partial_t^{(1)}\rho \bm{u} + \bm{u}\otimes\partial_t^{(1)}\rho \bm{u}^{\dagger} \\ - \bm{u}\otimes\bm{u}\partial_t^{(1)}\rho + \partial_t^{(1)}p\bm{I},
    \end{multline}
where we use,
    \begin{equation}
        \bm{u}\otimes\partial_t^{(1)}\rho \bm{u} = -\bm{u}\times\left[\bm{\nabla}\cdot\rho\bm{u}\otimes\bm{u} + \bm{\nabla}p\right],
    \end{equation}
and,
    \begin{equation}
        \bm{u}\otimes\bm{u}\partial_t^{(1)}\rho = - \bm{u}\otimes\bm{u} \bm{\nabla}\cdot\rho \bm{u},
    \end{equation}
to arrive at
    \begin{multline}
        \partial_t^{(1)}\left(\rho\bm{u}\otimes\bm{u} + p\bm{I}\right) = -\bm{\nabla}\cdot\rho\bm{u}\otimes\bm{u}\otimes\bm{u} -\bm{u}\otimes\bm{\nabla}p  \\ - {\left(\bm{u}\otimes\bm{\nabla}p\right)}^\dagger +  \partial_t^{(1)}p\bm{I},
    \end{multline}
where one can use Eq.~\eqref{eq:pressure_euler_CE} to get to,
    \begin{multline}
        \int \bm{v}\otimes\bm{v} f^{(1)}d\bm{v} =\\ -p\tau\left[ \left( \bm{\nabla}\bm{u} + \bm{\nabla}\bm{u}^{\dagger}\right) - \frac{r}{c_v}\bm{\nabla}\cdot\bm{u}\bm{I}\right].
    \end{multline}
Plugging this final expression into the momentum balance equation at order $\epsilon^2$,
    \begin{equation}
        \partial_t^{(2)}\rho \bm{u} + \bm{\nabla}\cdot\bm{T}_{\rm NS}  = 0,
    \end{equation}
where
    \begin{equation}
        \bm{T}_{\rm NS} = \tau p\left[\bm{\nabla}\bm{u} + \bm{\nabla}\bm{u}^\dagger - \frac{2}{D}\bm{\nabla}\cdot\bm{u}\bm{I}\right] + \tau p \left(\frac{2}{D}-\frac{r}{c_v}\right)\bm{\nabla}\cdot\bm{u}\bm{I}.
    \end{equation}
Moving on to the energy balance equation at order $\epsilon^2$, for the first split,
\begin{equation}
    \partial_t^{(2)} E + \bm{\nabla}\cdot\left(\int_{\mathbb{R}^D}\bm{v}g^{(1)}d\bm{v}\right) = 0,
\end{equation}    
where
\begin{multline}\label{eq:CE_noneq_q}
    \int_{\mathbb{R}^D}\bm{v}g^{(1)}d\bm{v} = -\tau\left[\partial_t^{(1)}\left(p+E\right)\bm{u} + \bm{\nabla}\cdot\left(\frac{p}{\rho}\left(E + p\right) \right. \right. \\ \left. \left. + p\bm{u}\times\bm{u} + \left(p + E\right)\bm{u}\times\bm{u}\right) \right].
\end{multline}
To expand this equation, we multiply the pressure balance equation at Euler level by $\bm{u}$ and obtain the following balance equation using momentum and continuity balance,
\begin{equation}
    \partial_t^{(1)}p\bm{u} + \bm{\nabla}\cdot p \bm{u}\otimes\bm{u} + \frac{p}{\rho}\bm{\nabla}p + \frac{r}{c_v}p\bm{u}\bm{\nabla}\cdot\bm{u} = 0.
\end{equation}
Using the same approach we can also derive the following additional balance equation,
\begin{equation}
    \partial_t^{(1)}E \bm{u} + \bm{\nabla}\cdot\left(E + p\right)\bm{u}\otimes\bm{u} + \frac{E}{\rho}\bm{\nabla}p - p\bm{u}\cdot\bm{\nabla}\bm{u} = 0.
\end{equation}
Plugging these equations into Eq.~\eqref{eq:CE_noneq_q} and after some algebra,
\begin{equation}\label{eq:vg1_eq}
    \int_{\mathbb{R}^D}\bm{v}g^{(1)}d\bm{v} = -\tau p \left(c_v+r\right)\bm{\nabla}T - \bm{u}\cdot\bm{T}_{\rm NS},
\end{equation}
which leads to:
\begin{equation}\label{eq:g_eps_2}
    \partial_t^{(2)}E - \bm{\nabla}\cdot\left[\tau p \left(c_v+r\right)\bm{\nabla}T\right] + \bm{\nabla}\cdot\left(\bm{u}\cdot\bm{T}_{\rm NS}\right) = 0.
\end{equation}
For the second splitting approach integrating Eq.~\eqref{Eq:CE_Eq_orders_2},
\begin{equation}
    \partial_t^{(2)}U + \bm{\nabla}\cdot\left[\int_{\mathbb{R}^D}\bm{v}g^{(1)}d\bm{v}\right] = \int_{\mathbb{R}^D} \mathcal{G}^{(2)}d\bm{v},
\end{equation}
where using,
\begin{equation}
    f^{(1)} = -\tau\left(\partial_t^{(1)}f^{(0)} +  \bm{v}\cdot\bm{\nabla}f^{(0)}\right),
\end{equation}
and,
\begin{equation}
    g^{(1)} = -\tau\left(\partial_t^{(1)}g^{(0)} +  \bm{v}\cdot\bm{\nabla}g^{(0)} - \mathcal{G}^{(1)}\right),
\end{equation}
one arrives at,
\begin{equation}
    \partial_t^{(2)} U - \bm{\nabla}\cdot \left[\tau p(c_v + r)\bm{\nabla}T\right] + \bm{T}_{\rm NS}:\bm{\nabla}\bm{u} = 0.
\end{equation}
For the third splitting approach,
\begin{equation}
    \partial_t^{(2)}E + \bm{\nabla}\cdot\left[\int_{\mathbb{R}^D}\bm{v}\left(g^{(1)} + \frac{\bm{v}^2}{2}f^{(1)}\right)d\bm{v}\right] = 0,
\end{equation}
where using,
\begin{equation}
    \int_{\mathbb{R}^D} \bm{v}\left(g^{(0)} + \frac{\bm{v}^2}{2}f^{(0)}\right)d\bm{v} = \left(E + p\right)\bm{u},
\end{equation}
and,
\begin{multline}
    \int_{\mathbb{R}^D} \bm{v}\otimes\bm{v}\left(g^{(0)} + \frac{\bm{v}^2}{2}f^{(0)}\right)d\bm{v} = \frac{p}{\rho}\left(E + p\right) + p\bm{u}\otimes\bm{u} \\ + \left(p + E\right)\bm{u}\otimes\bm{u},
\end{multline}
the same equation as in Eq.~\eqref{eq:vg1_eq} is recovered, which then leads to the energy balance of Eq.~\eqref{eq:g_eps_2}.
\end{document}